\documentstyle[12pt,epsf,aps]{revtex}

\def\prl#1#2#3{{ Phys. Rev. Lett.} {\bf #1}, #2 (#3)}

\def\pra#1#2#3{Phys. Rev. A {\bf #1}, #2 (#3)}
\def\prb#1#2#3{Phys. Rev. B {\bf #1}, #2 (#3)}
\def\pre#1#2#3{Phys. Rev. E {\bf #1}, #2 (#3)}

\def\physa#1#2#3{Physica A {\bf #1}, #2 (#3)}
\def\physd#1#2#3{Physica D {\bf #1}, #2 (#3)}

\def\rmp#1#2#3{Rev. Mod. Phys. {\bf #1}, #2 (#3)}

\def\siam#1#2#3{SIAM J. Appl. Math {\bf #1}, #2 (#3)}

\def\beq{\begin{equation}}
\def\bc{\begin{center}}
\def\ec{\end{center}}
\def\eqn{\end{equation}}

\begin{document}
\title{Nonequilibrium Dynamics in the Complex Ginzburg-Landau Equation}

\author{Sanjay Puri$^1$, Subir K. Das$^1$ and M.C. Cross$^2$} 
\address{$^1$School of Physical Sciences, Jawaharlal Nehru University \\
New Delhi -- 110067, India. \\
$^2$Department of Physics, California Institute of Technology \\
Pasadena, California 91125, U.S.A.\\}
\maketitle
\vspace{4cm}
\begin{abstract}
We present results from a comprehensive analytical and 
numerical study of nonequilibrium dynamics in the 2-dimensional complex
Ginzburg-Landau (CGL) equation. In particular, we use spiral defects
to characterize the domain growth law and the evolution
morphology. An asymptotic analysis of the single-spiral correlation
function shows a sequence of singularities -- analogous to those seen
for time-dependent Ginzburg-Landau (TDGL) models with $O(n)$ symmetry,
where $n$ is even.
\end{abstract}

\newpage
Much recent interest has focused on pattern formation in the complex
Ginzburg-Landau (CGL) equation:
\begin{equation}
\label{cgl}
\frac{\partial\psi(\vec r,t)}{\partial t}=\psi(\vec r,t)+
(1+i\alpha)\nabla^2\psi(\vec r,t)-(1+i\beta)|\psi(\vec r,t)|^2\psi(\vec
r,t) ,
\end{equation} 
where $\psi(\vec r,t)$ is a complex order-parameter field which depends
on space ($\vec r$) and time ($t$). In Eq. (\ref{cgl}), $\alpha$ and
$\beta$ are real parameters. The CGL equation arises in a range of
diverse contexts, as reviewed by Cross and Hohenberg \cite{ch}. This
universality arises from the fact that the CGL equation provides a
generic description of oscillations in a
spatially-extended system near a Hopf bifurcation \cite{ks}.

The CGL equation exhibits rich dynamical behavior with variation of the
parameters $\alpha$ and $\beta$, and the ``phase diagram'' has been
investigated (mostly numerically) in various studies \cite{cm}. In a
large range of parameter space, the emergence and interaction of spiral
defect structures characterizes the morphology. In this letter, we
study the nonequilibrium dynamics of the CGL equation resulting from a
small-amplitude random initial condition. In general, this
nonequilibrium evolution is referred to as ``phase ordering dynamics'' 
or ``domain growth'', and constitutes a well-studied example of
far-from-equilibrium statistical physics \cite{bin,bra}. Our analytical
understanding of phase ordering systems has depended critically upon
modeling the dynamics of defects in these systems (e.g., interfaces,
vortices, monopoles, etc.) \cite{bra}-\cite{maz}. In this letter, we
use spiral defect structures to characterize the
evolution morphology in the CGL equation.
Many novel features emerge in our study, which should
be of great relevance for both experiments and subsequent
numerical simulations.

For simplicity, we will focus on the CGL equation with $\alpha=0$ and
dimensionality $d=2$. However, the results presented here are also 
relevant for the cases with $\alpha\neq 0$ and $d>2$, as the 
underlying paradigm does not change, i.e., spirals continue to
determine the morphology in large regions of parameter space. Following
the work of Hagan \cite{hag}, Aranson {\it et al.} \cite{aakw}, and Chate and
Manneville \cite{cm}, let us briefly discuss the phase diagram of the $d=2$
CGL equation with $\alpha=0$. The limiting case $\beta=0$ corresponds to
the dynamical XY model, which is well understood. The appropriate (point)
defects are vortices, and domain growth is driven by the attraction and
annihilation of vortex-antivortex pairs. The relevant growth law for
the characteristic length scale is
$L(t) \sim (t/\ln t)^{1/2}$ \cite{pfg,bra}; and the analytic form of the
time-dependent correlation function (which characterizes the evolving
morphology) has been obtained by Bray and Puri, and (independently) Toyoki
\cite{bpt}. Without loss of generality, we focus on the case with
$\beta \geq 0$. For $0\leq\beta\leq\beta_{1}$ 
($\beta_{1}\simeq 1.397$ \cite{hag}),
spirals (which are asymptotically plane-waves) are linearly stable to
fluctuations. For $\beta_{1} < \beta\leq\beta_{2}$ ($\beta_{2}\simeq
1.82$ \cite{aakw,cm}), spirals are linearly unstable to 
fluctuations, but the growing
fluctuations are advected away, i.e., the spiral structure is globally
stable. Finally, for $\beta_{2}<\beta$, the spirals are globally unstable
and cannot exist for extended times \cite{aakw}. Our results
correspond to the parameter regime with $\beta\leq\beta_{2}$.

Figure $1$ shows the typical evolution from a small-amplitude random
initial condition for the case with $\beta=0.75$. Our numerical
simulations were performed by implementing an isotropic
Euler-discretization of Eq. (\ref{cgl}) on $N^2$-lattices ($N=256$ for
Figure 1), with periodic boundary conditions in both directions. The
discretization mesh sizes were $\Delta t = 0.01$ and $\Delta x = 1.0$.
In Figure $1$, we plot constant-phase regions and the 
relevant color-coding is
provided in the figure caption. The evolving morphology is
characterized by spirals and antispirals, and there is a typical length
scale $L$, e.g., inter-spiral spacing or the square root of 
inverse defect density, which is the definition we will use 
subsequently.

Figure 2(a) plots $\ln[L(t)]$ vs. $\ln t$ for 5 representative
values of $\beta$. The length-scale data was obtained from 5
independent runs on $N^2$-lattices with $N=1024$.
After an initial transient period, the length scale $L(t)$ 
should saturate to an equilibrium value ($L_s$) because of an effective
spiral-antispiral repulsive potential \cite{ch}. This should be
contrasted with the $\beta = 0$ case, where vortices continue to anneal
(at zero temperature) as $t \rightarrow \infty$. As a matter of fact, the
data for $\beta = 0.25, 0.5$ in Figure 2(a) does not exhibit this
morphological freezing on the time-scales of our simulation, though
signs of the crossover are evident for $\beta = 0.5$.

To understand this crossover behavior, we recall the analytical
solution for an $m$-armed spiral due to Hagan \cite{hag}:
\begin{equation}
\label{hgs}
\psi(\vec r,t)=\rho(r)\exp\left[-i\omega t+im\theta-i\phi(r)\right] ,
\end{equation}
where $\vec r\equiv (r,\theta)$; and $\omega=\beta(1-q^2)$, where 
$q$ is a constant which is determined by $\beta$ \cite{hag}.
The limiting forms of the functions $\rho(r)$ and $\phi(r)$ are
\begin{eqnarray}
\label{rp}
& &\rho(r) \rightarrow \sqrt{1-q^{2}}, ~~~\phi^{\prime}(r)\rightarrow q, 
\hspace{0.9cm}\mbox {as} \hspace{0.5cm} r\rightarrow \infty, \nonumber\\
& &\rho(r)\rightarrow ar^{m}, ~~~\phi^{\prime}(r)\rightarrow r, 
\hspace{0.5cm}~~~~~~~~~\mbox{as}\hspace{0.5cm}r\rightarrow 0,
\end{eqnarray}
where $a$ is a constant which is determined by finiteness conditions.
We will focus on the case with $m=\pm 1$, as only $1$-armed spirals
are stable in the evolution \cite{hag}. Furthermore, we are only
interested in distances $r\gg\xi$, where $\xi$ is the defect core. Thus,
we consider the spiral form in Eq. (\ref{hgs}) with
$\rho(r)=\sqrt{1-q^2}$ and $\phi(r)=qr$ (appropriate for $r \rightarrow
\infty$).

We expect that spirals behave similarly to vortices for $L<L_c$, where
$qL_c \sim O(1)$. Thus, the early evolution should be analogous to that
for the XY model, both in terms of the domain growth law and
correlation function. In Figure 2(a), the solid line has a slope of
$1/2$ and the initial growth (at least for $\beta \leq 0.75$) appears
to be consistent with the behavior for the XY model, i.e., $L(t) \sim
(t/\ln t)^{1/2}$ for $d=2$. We also expect the saturation length $L_s$
to scale with $L_c$. Figure 2(b) plots $L_{s}$ vs. $q^{-1}$ for a
range of $\beta$-values, and demonstrates that our numerical data
is consistent with $L_s \sim q^{-1}$.
We can also obtain the scaling law for the crossover time and
the corresponding numerical results (not shown here) are in agreement
with it.

Next, we consider the correlation function for the 
evolution morphology shown in
Figure $1$. It is obviously relevant to first consider the correlation
function for a single spiral of length $L$, as the snapshots in Figure
$1$ can be thought of as consisting of disjoint spirals of size $L$.
(Of course, this ignores modulations of the order parameter at
spiral-spiral boundaries but we will discuss those later.) We have
approximated the $1$-armed single-spiral solution as
$\psi(\vec r,t)\simeq \sqrt{1-q^2} \exp\left[-i\omega t+i(\theta-qr)\right]$.
The correlation function is obtained by considering the correlation
between points $\vec r_{1}$ and $\vec r_{2}~(= \vec r_{1}+\vec
r_{12})$ and integrating over $\vec r_{1}$ as follows:
\begin{eqnarray}
C(r_{12})&=&\frac{1}{V}\int d\vec r_{1} \mbox{Re}\left\{\psi(\vec
r_{1},t)\psi(\vec r_{2},t)^{*}\right\}h(L-r_{2}) \nonumber\\
&=&\frac{(1-q^2)}{V} \mbox{Re}\int d\vec r_{1}
\exp\left[i(\theta_{1}-\theta_{2}-qr_{1}+q|\vec r_{1}+\vec
r_{12}|)\right]h(L-|\vec r_{1}+\vec r_{12}|),
\end{eqnarray}
where $V$ is the spiral volume; and we have
introduced the step function $h(x)=1~(0)$ if $x \geq 0~(x<0)$. The
step function ensures that we do not include points which 
lie outside the defect of size $L$.

It is convenient to introduce variables
$\theta_{1}-\theta_{12}=\theta$; $x=r_{1}/L$; $r=r_{12}/L$, to obtain
\begin{eqnarray}
\label{cr}
C(r_{12})=\frac{(1-q^2)}{\pi}\mbox{Re}\int_{0}^{1}dx x\int_{0}^{2\pi}d\theta
\frac{x+re^{i\theta}}
{(x^{2}+r^{2}+2xr\cos\theta)^{1/2}}\times\nonumber\\
\exp \left[ -iqL\left\{ x-(x^{2}+r^{2}+2xr\cos\theta)^{1/2} \right\} \right]
h[1-(x^2+r^2+2xr \cos\theta)^{1/2}],
\end{eqnarray}
where we have used $V=\pi L^2$ in $d=2$.
Thus, the scaling form of the single-spiral correlation function is
$C(r_{12})/C(0)\equiv g(r_{12}/L,q^{2}L^{2})$. In general, 
there is no scaling with the
spiral size because of the additional factor $qL$. We recover scaling
only in the limit $q=0$ ($\beta=0$), which corresponds to the case of a
vortex. Essentially, spirals of different sizes are not
morphologically equivalent because there is more rotation in the phase
as one goes out further from the core. 

Figure 3 plots $C(r_{12})/C(0)$ vs. $r_{12}/L$ for
the case with $\beta=0.75$ ($q\simeq 0.203$).
These results are obtained by a direct numerical integration of
Eq. (\ref{cr}). We consider $4$ different values of $L$. The functional
form in Figure $3$ exhibits near-monotonic behavior for small values
of $L$ (i.e., in the vortex or XY limit); and pronounced oscillatory 
behavior for larger
values of $L$, as is expected from the integral expression. Notice that
$r_{12}/L\leq 2$ -- larger values of $r_{12}$ correspond to the point
$\vec r_{2}$ lying outside the defect.

The asymptotic behavior of the correlation function in the limit
$r=r_{12}/L \rightarrow 0$ (though $r_{12}/\xi\gg 1$) is of
considerable importance as it determines the tail of the momentum-space
structure factor \cite{bra}. In particular, we are interested in the
singular part of the correlation function as $r \rightarrow 0$. In this
limit, we can discard the step function in Eq. (\ref{cr}) as it only
provides corrections at the edge of the defect. The asymptotic analysis
of the integral in Eq. (\ref{cr}) involves considerable algebra, which
we will report in detail elsewhere. Here, we confine ourselves to
quoting the final result for the singular part of $C(r_{12})$:
\begin{eqnarray}
\label{csing}
C_{\mbox{\scriptsize sing}}(r_{12})&=&
\frac{1}{2}\sum_{p=0}^{\infty}\sum_{m=0}^{\infty}(-1)^{p+m}
\frac{(qL)^{2(p+m)}}{(2p)!(2m)!}\frac{\Gamma\left(\frac{1}{2}+m\right)^{2}}
{\Gamma\left(\frac{1}{2}-p\right)^{2}(m+p+1)!^{2}}\times\nonumber\\
& & ~~~~~~~~~~~~~~~~~~~(2m+1)(2p+1)r^{2(m+p+1)}\ln r .
\end{eqnarray}

Eq. (\ref{csing}) is one of the central results
of this paper and we would like
to briefly discuss its implications. The leading-order singularity is
the same as that for the XY model ($\beta=q=0$),
$C_{\mbox{\scriptsize sing}}(r_{12})=\frac{1}{2}r^{2}\ln r$ \cite{bh}, 
as expected. However, there is also a sequence of sub-dominant singularities
proportional to $(qL)^{2}r^{4}\ln r$,
$(qL)^{4}r^{6}\ln r$, etc., and these become increasingly important as the
length scale $L$ increases. These sub-dominant terms in
$C_{\mbox{\scriptsize sing}}(r_{12})$ are reminiscent of the leading-order
singularities in models with $O(n)$ symmetry, where $n$ is even 
\cite{bra,bh}.
Of course, in the context of $O(n)$ models, these singularities only
arise for $n\leq d$ as there are no topological defects unless this
condition is satisfied. In the present context, all these terms are
already present for $d=2$. The implication for
the structure-factor tail is a sequence of power-law decays with
$S(k)\sim (qL)^{2(m-1)}L^{d}/(kL)^{d+2m}$, where $m=1,2$, etc. 
Thus, though the true asymptotic behavior in $d=2$ 
is still the generalized Porod tail, $S(k) \sim L^2 (kL)^{-4}$, it may be
difficult to disentangle this from other power-law decays.

Finally, Figure $4$ compares our numerical data for the correlation
function with the functional form of the single-spiral correlation function.
Recall that the correlation function does not scale with the
characteristic length because of the spiral nature of the defects. 
In Figures 4(a)-(c), we have plotted 
numerical data for $C(r_{12},t)/C(0,t)$ vs.
$r_{12}$ at $t=500$, and $\beta=0.75, 1.0, 1.25$. For the comparison
with Eq. (\ref{cr}), the length scale $L$ is taken to be an 
adjustable parameter. In each case, the best-fit value of $L$ matches
the length scale obtained from the inverse defect density
(see Figure 2(a)) within 10 percent. As is seen from Figure $4$,
the single-spiral correlation function is in good agreement with the
numerical data for the multi-spiral morphology upto (approximately) the
first minimum. As a matter of fact, the agreement is excellent (perhaps
fortuitously) for $\beta=1.25$, shown in Figure 4(c).

In the context of phase ordering dynamics, the Gaussian auxiliary field
(GAF) ansatz \cite{bra}-\cite{maz} has proven particularly useful for the
characterization of multi-defect morphologies. We have critically
examined the utility of the GAF ansatz in the present context \cite{dpc}
and find that it is only reasonable at early times -- where, in any 
case, the ordering process is analogous to that for the XY model. We are
presently studying methods of improving the GAF ansatz for the CGL
equation and will discuss this elsewhere.

More generally, the utility of the GAF ansatz arises from the
summation over phases from many defects, which results in a near-Gaussian
distribution for the auxiliary field. However, in the present context, 
the shocks between spirals effectively isolate one spiral
region from the influence of other regions. As a matter of fact, the
waves from other spirals decay exponentially through the shock and the
phase of a point is always dominated by the nearest spiral. Therefore,
we expect that the correlation function will be dominated by the
single-spiral result -- in accordance with our numerical results.

To summarize: we have undertaken a detailed analytical and numerical
study of nonequilibrium dynamics in the CGL equation. For early times
($L<L_{c}\sim q^{-1}$), the domain growth process is analogous to that
for the XY model, which is well understood. At later times, distinct
effects due to spirals are seen and the evolving system freezes (in a
statistical sense) into a multi-spiral morphology. We have undertaken
an asymptotic analysis of the correlation function $C(r_{12})$ for a
single spiral. It exhibits a sequence of singularities as $r_{12}/L
\rightarrow 0$. Furthermore, this correlation function is in good
agreement with the numerical data for multi-spiral morphologies, over
an extended range of distances.

\section*{Acknowledgements}

SP is grateful to A.J. Bray and H. Chate for useful discussions. SKD
is grateful to the University Grants Commission, India, for financial
support in the form of a research fellowship. MCC thanks the School 
of Physical Sciences, JNU for hospitality during the stay in 
which this work was begun.

\newpage

\section*{Figure Captions}

\vskip0.5cm
\noindent{\bf Figure 1:} Evolution of the CGL equation from a 
small-amplitude random initial condition. The evolution
pictures were obtained from an Euler-discretized version 
of Eq. (\ref{cgl}) with $\alpha=0,~\beta=0.75$, 
implemented on an $N^{2}$-lattice ($N=256$). The
discretization mesh sizes were $\Delta t=0.01$, $\Delta x=1.0$; and
periodic boundary conditions were imposed in both directions. The
snapshots show regions of constant phase $\theta_{\psi}=\tan^{-1}
(\mbox{Im}\psi/\mbox{Re}\psi)$, measured in radians, with the
following color coding: $\theta_{\psi} \in [1.85, 2.15]$ (black); 
$\theta_{\psi} \in [3.85, 4.15]$ (red); $\theta_{\psi} \in [5.85, 6.15]$ 
(green). The snapshots are labeled by the appropriate evolution times. \\
\ \\
{\bf Figure 2:} (a) Plot of $\ln[L(t)]$ vs. $\ln t$ for $\alpha=0$ and
$\beta = 0.25, 0.5, 0.75, 1.0, 1.25$ -- denoted by the specified symbols. 
The characteristic length scale, $L(t)$, is obtained from the square 
root of the inverse defect density -- measured directly from snapshots
as shown in Figure $1$. The numerical data shown here was obtained as an
average over $5$ independent runs for $N^2$-lattices (with $N=1024$).
The solid line has a slope of $1/2$.\\
(b) Plot of saturation length $L_s$ vs. $q^{-1}$ for a range of 
$\beta$-values. The corresponding values of $q$ 
(as a function of $\beta$) are obtained from
Hagan's solution, cf. Figure $5$ of Ref. \cite{hag}. The solid line
denotes the best linear fit to the numerical data.\\
\ \\
{\bf Figure 3:} Correlation function for the $1$-armed spiral
solution when $\beta=0.75$ ($q \simeq 0.203$). We plot $C(r_{12})/C(0)$ vs.
$r_{12}/L$ for different spiral sizes, $L=15,25,50,100$ -- denoted by the
specified line-types. These results are obtained from a direct numerical
integration of Eq. (\ref{cr}). \\
\ \\
{\bf Figure 4:} Numerical data for the correlation function
$C(r_{12},t)/C(0,t)$ vs. $r_{12}$ at $t=500$
for the cases $\alpha=0$ and (a) $\beta=0.75$; (b) $\beta=1.0$; 
(c) $\beta=1.25$. The numerical data
was obtained as an average over $5$ independent runs for $N^2$-lattices 
(with $N=1024$). The solid line refers to the numerical
integration of Eq. (\ref{cr}) with $L$ as an adjustable parameter.
Subsequently, the $r_{12}$-axis is scaled so that the point
$C(r_{12},t)/C(0,t)=1/2$ is matched for the numerical data and the
analytical expression.\\

\end{document}